\documentclass[apj,useAMS,usenatbib]{aastex}
\usepackage{times}
\usepackage{amssymb}
\usepackage{epsfig}
\usepackage{rotating}
\usepackage{lscape}

\def\ga{\mathrel{\raise0.35ex\hbox{$\scriptstyle >$}\kern-0.6em
\lower0.40ex\hbox{{$\scriptstyle \sim$}}}}
\def\la{\mathrel{\raise0.35ex\hbox{$\scriptstyle <$}\kern-0.6em
\lower0.40ex\hbox{{$\scriptstyle \sim$}}}}
\def\co{CO~{\it J}=1-0 }
\def\cotwo{CO~{\it J}=2-1 }

\def\loj{low-{\it J}~}

\newcommand{\kms}{km~s$^{-1}$}

\shorttitle{Molecular gas in cluster galaxies at $z \sim 1$}
\shortauthors{Wagg et al.}
\slugcomment{accepted for publication in ApJ}

\begin{document}

\title{CO \textit{J}=2-1 line emission in cluster galaxies at $z \sim 1$: fueling star formation in dense environments\footnote{Based on observations carried out with the IRAM Plateau de Bure Interferometer. IRAM is supported by INSU/CNRS (France), MPG (Germany) and IGN (Spain).}}

\author{
Jeff~Wagg$^{1}$, Alexandra~Pope$^{2}$, Stacey~Alberts$^{2}$, Lee~Armus$^{3}$, Mark~Brodwin$^{4}$, Robert~S.~Bussmann$^{5}$, Vandana~Desai$^{3}$, Arjun~Dey$^{6}$, Buell~Jannuzi$^{6}$, Emeric~Le~Floc'h$^{7}$, Jason~Melbourne$^{8}$, and Daniel~Stern$^{9}$}

\affil{$^{1}$European Southern Observatory, Casilla 19001, Santiago, Chile}\email{jwagg@eso.org}

\affil{$^{2}$Department of Astronomy, University of Massachusetts, Amherst, MA 01003, USA}

\affil{$^{3}$Spitzer Science Center, California Institute of Technology, MS 220-6, Pasadena, CA 91125, USA}

\affil{$^{4}$Department of Physics, University of Missouri, 5110 Rockhill Road, Kansas City, MO 64110}

\affil{$^{5}$Harvard-Smithsonian Center for Astrophysics, 60 Garden Street, Cambridge, MA 02138, USA}

\affil{$^{6}$National Optical Astronomy Observatory, Tucson, AZ 85726-6732, USA}

\affil{$^{7}$AIM, CNRS, Universit\'e Paris Diderot, B\^at. 709, CEA-Saclay, 91191 Gif-sur-Yvette Cedex, France}

\affil{$^{8}$California Institute of Technology, Pasadena, CA 91125, USA}

\affil{$^{9}$Jet Propulsion Laboratory, California Institute of Technology, Pasadena, CA 91109, USA }

\begin{abstract}
 We present observations of \cotwo line emission in infrared-luminous cluster galaxies at $z \sim 1$ using the IRAM Plateau de Bure Interferometer. Our two primary targets are optically faint, dust-obscured galaxies (DOGs) found to lie within 
  2$\,$Mpc of the centers of two massive ( $>10^{14}$~M$_{\odot}$) galaxy clusters. CO line emission is not detected in either DOG. We calculate 3-$\sigma$ upper limits to the \cotwo line luminosities, $L^\prime_{\rm CO}< 6.08\times 10^{9}$ and $< 6.63\times 10^{9}$~K~km~s$^{-1}$~pc$^{2}$. Assuming a CO-to-H$_2$ conversion factor derived for ultraluminous infrared galaxies in the local Universe, this translates to limits on the cold molecular gas mass of $M_{\rm H_2} < 4.86 \times 10^{9} \:$~M$_{\odot}$ and  $M_{\rm H_2} < 5.30 \times 10^{9} \:$~M$_{\odot}$. Both DOGs exhibit mid-infrared continuum emission that follows a power-law, suggesting that an AGN contributes to the dust heating. As such, estimates of the star formation efficiencies in these DOGs are uncertain.
A third cluster member with an infrared luminosity, $L_{\rm IR} < 7.4\times 10^{11}$~L$_{\odot}$, is serendipitously detected in \cotwo line emission in the field of one of the DOGs located roughly two virial radii away from the cluster center. The optical spectrum of this object suggests that it is likely an obscured AGN, and the measured CO line luminosity is $L^\prime_{\rm CO}= (1.94 \pm 0.35)\times 10^{10}$~K~km~s$^{-1}$~pc$^{2}$, which leads to an estimated cold molecular gas mass $M_{\rm H_2} = (1.55 \pm 0.28)\times 10^{10}$~M$_{\odot}$. 
A significant reservoir of molecular gas in a $z \sim 1$ galaxy located away from the cluster center demonstrates that the fuel can exist to drive an increase in star-formation and AGN activity at the outskirts of high-redshift clusters.
\end{abstract}


\section{Introduction}

In the local Universe, the average star formation rate (SFR) in galaxies depends on the local environment, 
decreasing with increasing galaxy space density (e.g. Lewis et al.\ 2002; Kauffmann et al.
\ 2004). 
Recent studies have demonstrated that the 
average star formation rate in galaxies in rich environments is much higher
at $z \sim 1$ than it is locally (Elbaz et al. 
2007; Cooper et al. 2008), suggesting that galaxies in $z \ga 1$ clusters exhibit more intense star-formation than their analogues in the local Universe.
Indeed, the average SFR in galaxies is higher in massive clusters at these redshifts (Tran et al.~2010; Hilton et al.~2010; Brodwin et al. \textit{in prep}). Furthermore, the number of cluster galaxies with an active galactic nucleus (AGN) has been found to increase out to redshifts $z > 1$ (e.g. Galametz et al.\ 2009; Martini et al.\ 2009). At intermediate redshifts, recent studies have found that luminous infrared galaxies are preferentially forming stars at the outskirts of some massive clusters (e.g. Haines et al.\ 2010), suggesting that star-formation may be quenched within the central regions at these epochs.
 This star-formation and AGN activity in high-redshift
 cluster galaxies should be fueled by significant reservoirs of
 molecular gas, and observations of this gas can provide insight into the processes governing galaxy formation in dense environments.

The most effective
 means of studying the molecular gas in high-redshift galaxies is
 through observations of redshifted CO line emission (see reviews by
 Solomon \& Vanden~Bout 2005; Carilli et al.\ 2011a), as the luminosity in the lowest-\textit{J} CO line transitions (\textit{J}=2-1 or \textit{J}=1-0) provide an estimate of the total mass in cold molecular gas. 
Sensitive interferometric imaging of CO line emission redshifted to mm and cm-wavelengths has been used to probe the molecular gas content of members of galaxy overdensities at redshifts, $z > 2$ (e.g. Nesvadba et al.\ 2009; Riechers et al.\ 2010; Carilli et al.\ 2011b). These observations revealed that gas-rich galaxies do reside in proposed galaxy overdensities at high-redshift, presumably fueling starburst and AGN activity in objects that are rapidly evolving toward massive galaxies at $z \sim 0$. In the nearby Universe, most early-type galaxies inhabiting rich clusters  are gas-poor (e.g. Combes et al.\ 2007), as the cold gas has been consumed during star-formation, or removed through processes such as ram pressure stripping. At intermediate redshifts, \co line emission has been detected  in two luminous infrared galaxies (LIRGs) found within 2 to 3 virial radii of the center of the rich cluster Cl~0024+16 at $z = 0.395$ (Geach et al.\ 2009). It is possible that such systems will consume their gas rapidly as they enter the cluster. However, these clusters are still at relatively low redshift, in the regime where we expect star formation activity to be suppressed in the harsh cluster environments. To date, there have been no studies of the cold molecular gas traced by redshifted CO line emission in massive ($>$10$^{14}$~M$_{\odot}$) galaxy clusters at $z>1$.

\begin{figure*}[ht]
\centering
\epsfig{file=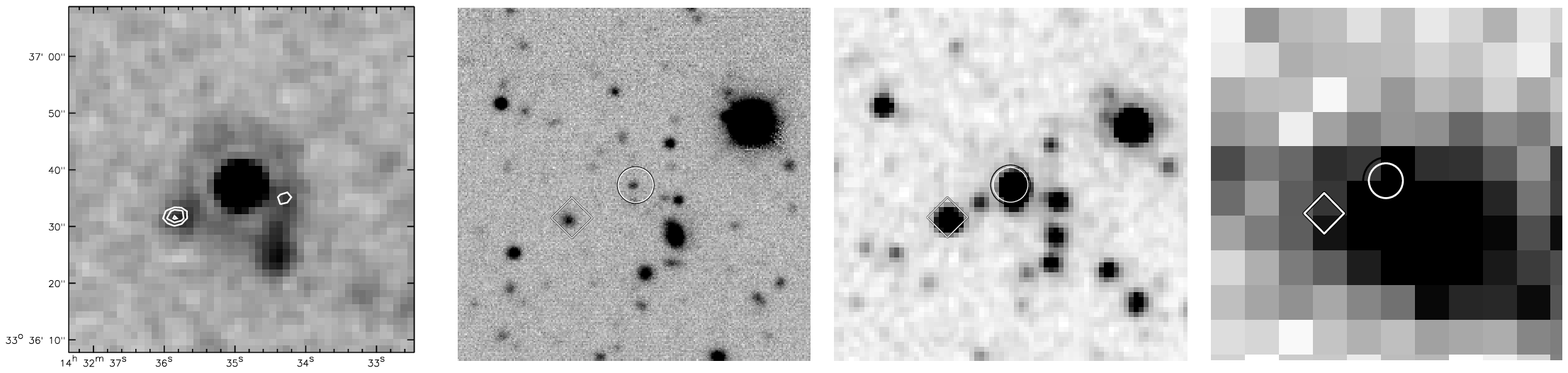,width=6.5in}

\epsfig{file=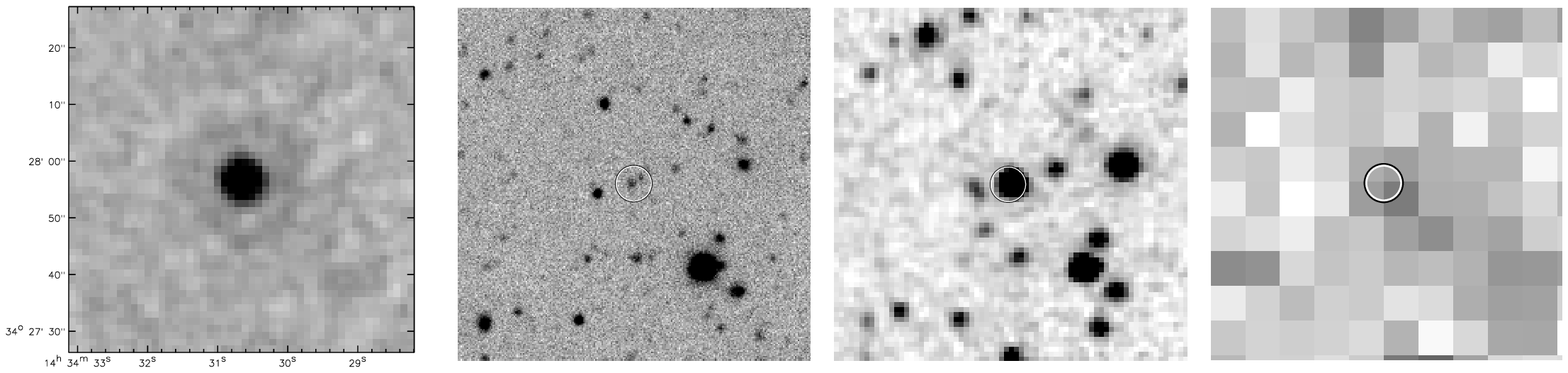,width=6.5in}
\caption{Images of the DOG1 (\textit{top}) and DOG2 (\textit{bottom}) fields. From left to right are shown $1' \times 1'$ images at 24~$\mu$m, $R-$band, 4.5~$\mu$m, and SPIRE 250~$\mu$m. The \textit{circles} denote the positions of the two DOGs studied here, while the \textit{diamond} indicates the position of the CO-detected source we refer to as COserendip. Overlaid on the 24$\mu$m image of the DOG1 and COserendip field are CO intensity contour levels of $(3, 4, 5) \times \sigma$, where $\sigma = 0.27$~mJy per 500~km/s channel.} 
\label{fig:stamps}
\end{figure*}

The formation of massive galaxies at high-redshift is often traced by strong infrared emission, which probes an active phase of dust obscured star formation and AGN activity, such as that seen in ultra-luminous infrared galaxies (ULIRGs) in the nearby Universe. Local and high redshift ULIRGs are invariably observed to be rich in molecular gas which provides fuel for ongoing star formation activity (Solomon \& Vanden Bout 2005).
One method used to identify ULIRGs at high-redshift is through a high mid-infrared-to-optical flux density ratio, which has been used to reveal a population of $z\sim 1 -3$ dust-obscured galaxies (DOGs; Dey et al.\ 2008; Fiore et al.\ 2008) which are strongly clustered (Brodwin et al.\ 2008). DOGs have high estimated IR luminosities, $L_{\rm IR} = $10$^{12-13}$~$L_\odot$, that are likely powered by a combination of star-formation and AGN activity (Dey et al.\ 2008; Desai et al.\ 2009), and theoretical models of merging galaxies are able to reproduce many of the observed properties of DOGs (Narayanan et al.\ 2010). 
The significant star-formation rates of starburst-dominated DOGs implied by their FIR luminosities ($SFR > 500$~M$_{\odot}$~yr$^{-1}$) are expected to be fueled by large masses of molecular gas.
 We find that some of the DOGs lie within 1--2$\,$Mpc of the centers of massive, high-redshift galaxy clusters from the IRAC Shallow Cluster Survey (ISCS, Eisenhardt et al.~2008). 
These clusters have masses $>10^{14}M_{\odot}$ (Brodwin et al.\ 2011; Jee et al.~2011), estimated from weak lensing and X-ray observations. The DOGs in these clusters provide excellent targets for studying the gas properties of infrared luminous galaxies residing in the densest environments at $z > 1$.

In the present work, we use the IRAM Plateau de Bure Interferometer (PdBI) to conduct a sensitive search for \cotwo line emission in two DOGs associated with massive galaxy clusters at $z > 1$ in order to measure their total cold molecular gas masses and study the star formation efficiency of galaxies in extreme cluster environments. Throughout this analysis, we adopt a cosmological 
model with $(\Omega_\Lambda, \Omega_m, h) = (0.73, 0.27, 0.71)$ (Spergel et al.\ 2007). At a redshift of $z = 1.2$, the angular scale is 8.346~kpc/$''$.

\section{Our targets}
\label{sec:sources}

Dey et al.\ (2008) identify 2603 DOGs in the NOAO Deep Wide Field Survey of the 9~square degree 
Bo\"otes field  (Jannuzi \& Dey 1999). This field contains a sample of over twenty massive galaxy clusters spectroscopically confirmed to lie at $z > 1$ (Stanford et al.\ 2005; Elston et al.\ 2006; Brodwin et al.\ 2006; Eisenhardt et al.\ 2008; Brodwin et al.\ 2011). Three DOGs with spectroscopic redshifts are found to lie within a distance of 2~Mpc from the centers of these known clusters. This separation places them close to their expected virial radii given the estimated masses of these clusters ($\ga$$10^{14}$~M$_{\odot}$, Jee et al.~2011). Of the three DOGs associated with clusters, the 24~$\mu$m flux densities predict that two of these should have infrared luminosities, $L_{\rm IR} \ga 10^{12}$~L$_{\odot}$, making them suitably bright for mm-wavelength observations of CO line emission. SST24~J143234.9+333637 (hereafter `DOG1') lies at a projected separation of $\sim$2~Mpc from the center of cluster ISCS~J1432.4+3332 ($r_{200} = 1.06^{+0.11}_{-0.09}$~Mpc), and SST24~J143430.6+342757 (hereafter `DOG2') is at a projected separation of $\sim$0.8~Mpc from the center of cluster ISCS J1434.5+3427 ($r_{200} = 0.82^{+0.19}_{-0.14}$~Mpc). Both DOG1 and DOG2 have optical spectroscopic redshifts (Table~\ref{tab:tab1}). All source names indicate  J2000 coordinates. The shape of the mid-infared spectral energy distributions of DOG1 and DOG2 can be described by a power-law, indicating that AGN are present (Stern et al.\ 2005; Dey et al.\ 2008). The optical Keck/LRIS spectra of DOG1 reveals that this object is a type-2 AGN at z=1.1160. Narrow,
high-ionization lines of Ne are clearly detected, as well as \ion{He}{2} and
\ion{C}{3}]. The Keck/DEIMOS  optical spectrum of DOG2 shows [\ion{O}{2}] emission at $z=1.2405$, with low signal-to-noise ratio detections of corresponding \ion{Ca}{2} HK absorption and a 4000\AA\ break.  The optical spectrum shows no high ionization lines indicative of an active galaxy. Independent of whether the AGN, or star-formation is the dominant source of their infrared luminosities, we expect the DOGs to have significant masses of cold molecular gas, as is typically found in objects whose dusty interstellar medium is strongly irradiated.

 Optical-to-submm wavelength images for each of our two target fields are shown in Figure~\ref{fig:stamps}. 
In order to better interpret the CO observations of our targets, we fit the available infrared photometry with spectral energy distribution (SED) templates to estimate their infrared (8-1000$\mu$m)  and far-infrared luminosities (40-500$\mu$m). Publicly available \textit{Herschel}/SPIRE data now exist for both targets in Bo\"otes from the HerMES survey (Oliver et al.~2010). We reprocessed the SPIRE maps to remove striping, astrometry offsets and glitches before performing our own photometry measurements on the images at 250, 350 and 500~$\mu$m. DOG2 is formally undetected in the SPIRE bands while DOG1 and a second 24~$\mu$m source (SST24~J143235.8+333632  - described in \S~\ref{sec:comember}) are both part of a group of 24~$\mu$m sources which are blended together as a single source in the lower resolution SPIRE maps, highlighted by the 250~$\mu$m image shown in Figure~1. We measure the SPIRE flux densities at the exact position of the 24~$\mu$m counterparts, noting that the resulting infrared luminosities should be considered upper limits, since there may be significant contribution from the nearby blended sources. We fit the photometry using the new starburst and AGN SED templates of Kirkpatrick et al.~(\textit{in preparation}) which are developed using a complete suite of data for high redshift galaxies including IRS spectroscopy, \textit{Spitzer} imaging from 3.6-to-70~$\mu$m, and \textit{Herschel} imaging from 100-to-500~$\mu$m. The resulting luminosities are listed in Table~1. Both DOG1 and DOG2 exhibit mid-infrared continuum emission that can be described by a power-law, suggesting that AGN are present.  However, the current data prohibit us from determining if star-formation or AGN activity dominate the far-infrared luminosities.

Using broad-band optical through near-IR photometry, we constrain the stellar masses of our targets.  High signal-to-noise detections at each of \textit{B$_W$}, \textit{R}, \textit{I}, \textit{J}, \textit{H}, and \textit{Ks}  
 bands (Jannuzi \& Dey 1999; Gonzalez et al.\ \textit{in prep.}), were fit with  synthesized stellar population models following the procedure described in Bussmann et al.\ (2011), where the best-fit stellar mass, $M_*$, and the 1$\sigma$ uncertainties are determined by marginalizing over stellar population age and reddening.  Detections of these objects in the IRAC bands (Ashby et al.\ 2009) were not used in the fitting. 
A simple stellar population is assumed, as well as a Chabrier initial mass function (Chabrier 2003), and the Bruzual \& Charlot (2003) synthesis models. The best-fit stellar masses and their 1-$\sigma$ uncertainties (for the given choice of star formation history and initial mass function) are listed in Table~1.  More complex star-formation histories can lead to an increase in these stellar mass estimates of up to $\sim 0.5$~dex (Bussmann et al. 2011).

\section{Observations and data analysis}
\label{sec:obs}

\subsection{PdBI observations of \cotwo line emission}

At the redshifts of DOG1 and DOG2, the 230.538~GHz \cotwo line is redshifted to 108.950 and 102.896~GHz, which is accessible with the 3~mm band receivers on the PdBI.  
Observations were made with five antennas of the PdBI interferometer on four dates between August 29 and Oct 3, 2010, while the array was in the compact 5Dq configuration. A total of 12.7~hours (including overheads) was spent observing DOG1, while another 10~hours was spent observing the DOG2 field. The correlator was set up to sample 3.6~GHz of bandwidth in two polarizations, covering 9,880 and 10,460~km~s$^{-1}$ at these frequencies. With such a large bandwidth and a field of view of $\sim$0.5~sq.~arcmin., we are also able to conduct a serendipitous search for \cotwo line emission  in additional cluster members. The lowest spectral resolution mode of the correlator was adopted, and the calibrated data were resampled to a resolution of 40.1~MHz. 1749+096 and 1424+366 were observed for complex gain calibration, while observations of MWC349 and 3C345 were used to calibrate the flux density and spectral bandpass. After these observations were made, a hardware problem was uncovered by the IRAM staff which would have affected the phases on antenna 5 baselines for observations taken during our October 3 observations of DOG2. After checking these data, we do not find any clear evidence for problems with the phases, which were likely negated due to the close proximity of our phase calibrator to the target field (1424+366; $\theta \sim $2.5$^\circ$).

Data were calibrated using the IRAM GILDAS\footnote{http://www.iram.fr/IRAMFR/GILDAS} \textit{CLIC} software package, while the imaging was performed with GILDAS \textit{mapping}. Natural weighting was used in the imaging, and the final synthesized beamsizes are $5\farcs 0 \times 3\farcs 8$ in the DOG1 image, and $4\farcs 9 \times 3\farcs 9$ in the DOG2 image. The image map sizes are $32\farcs0 \times 32\farcs0$, chosen to include a significant fraction of the field of view defined by the primary beam. The calibrated \textit{uv} data are also processed separately by a routine written by the IRAM staff that searches for emission features in the spectral line data cubes. We identify one source of significant emission (5.5-$\sigma$) at $\sim$12$''$ from the center of the DOG1 map, and this object is described  in \S~\ref{sec:comember}. Due to its distance from the center of the image, the flux density has been corrected for the primary beam by a factor of 1.033. We have verified that the detection remains after excluding a small subset of antenna 5 data that are potentially affected by a hardware problem.

\begin{landscape}
\begin{table}[ht]
\caption{Observed properties of $z\sim 1$ cluster galaxies.
\label{tab:tab1}}
\begin{center}
\begin{tabular}{ccccccccccc} 
\hline 
\hline 
Source  & $z $ &  $S_{24\mu m}$ & L$_{\rm IR}$ (8-1000$\mu$m) & L$_{\rm FIR}$ (40-500$\mu$m)  & $\log(M_*)$$^\dagger$  & $\nu_{{\rm CO} 2-1}$ & RMS$^\star$ &  $L^\prime_{\rm CO}$ \\
       &    &   [mJy]   &   [$L_{\odot}$]  &  [$L_{\odot}$]   & [M$_\odot$]  & [GHz ]  &  [mJy] &  [K~\kms~pc$^2$]  \\
       &          &                     &               &       &        \\
\hline 
DOG1  & 1.1160  & $2.92\pm 0.07$  &  $<2.8\times10^{12}$$^\Diamond$ & $<1.6\times10^{12}$$^\Diamond$ &        $10.57 \pm 0.14$ &  108.950 & 0.70 & $< 6.08\times 10^{9\ddagger}$ \\
DOG2 & 1.2405 &  $1.67\pm 0.05$ & $<1.7\times10^{12}$$^\Diamond$ & $<5.8\times10^{11}$$^\Diamond$ & 
$10.94 \pm 0.13$ &  102.896 & 0.62 & $< 6.63\times 10^{9\ddagger}$  \\
COserendip & 1.1147 (CO) & $0.45\pm 0.04$ &  $<7.4\times10^{11}$$^\Diamond$ & $<4.9\times10^{11}$$^\Diamond$  & $10.97 \pm 0.14$  &  109.018 & 0.66 & $ (1.94 \pm 0.35)\times 10^{10}$  \\      
\hline
\end{tabular}
\vskip 0.1in
\noindent $^\dagger${}The uncertainties on $M_*$ are for the specific choice of star formation history and initial mass function, and therefore underestimate the uncertainty in the absolute values. \\
\noindent $^\star${}The quoted RMS noise values are calculated for a velocity resolution of $100$~\kms. \\
\noindent $^\Diamond$ Upper-limits to the infrared and far-infrared luminosities are due to source blending, or non-detections in the SPIRE submm bands. \\
\noindent $^\ddagger${}3-$\sigma$ upper limits assuming a 300~km~s$^{-1}$ CO linewidth. \\
\end{center}
\end{table}
\end{landscape}

\section{Results}
\label{sec:results}

\subsection{\cotwo line emission in DOGs}

\begin{figure}[ht]
\centering
\epsfig{file=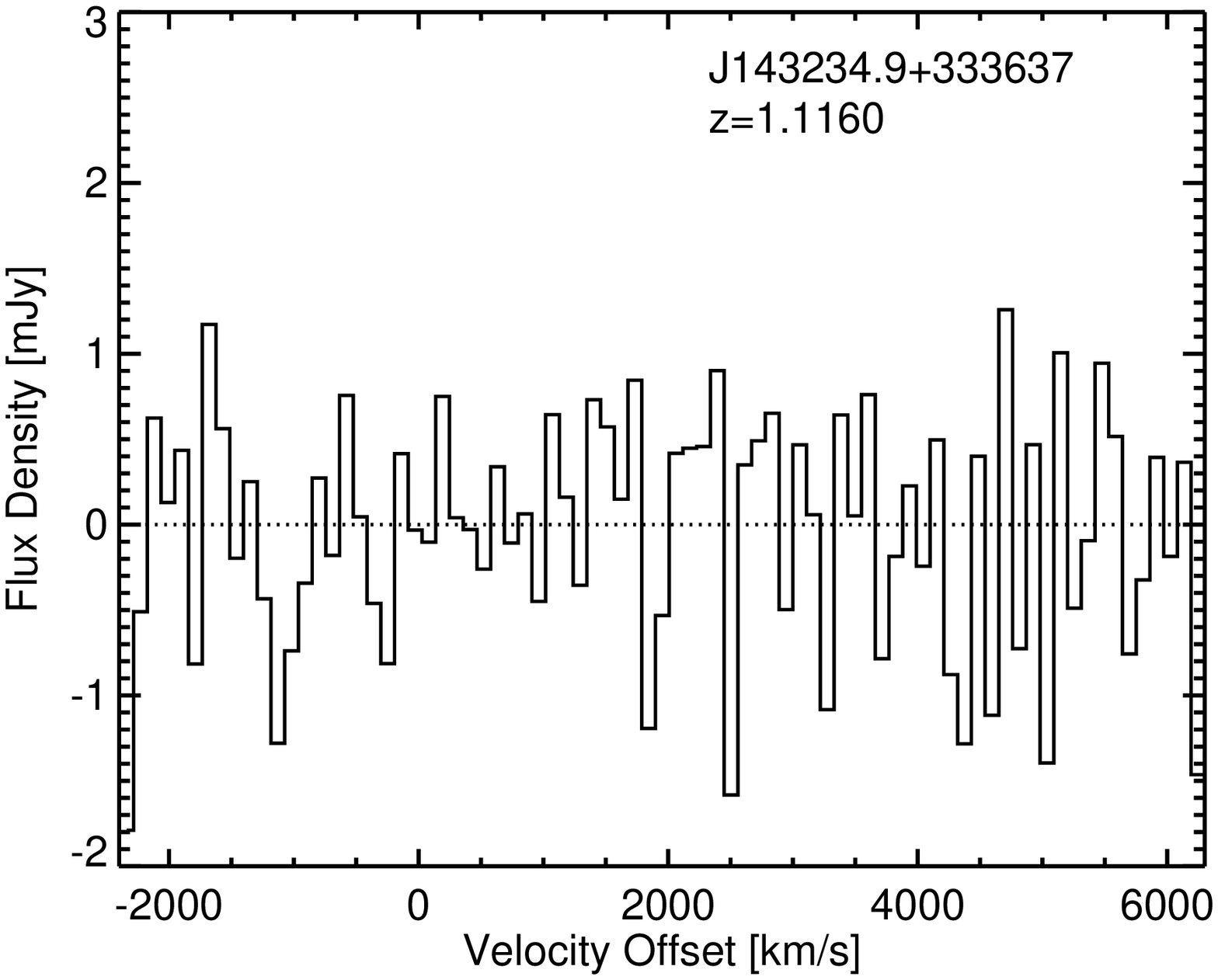,width=3.in}

\epsfig{file=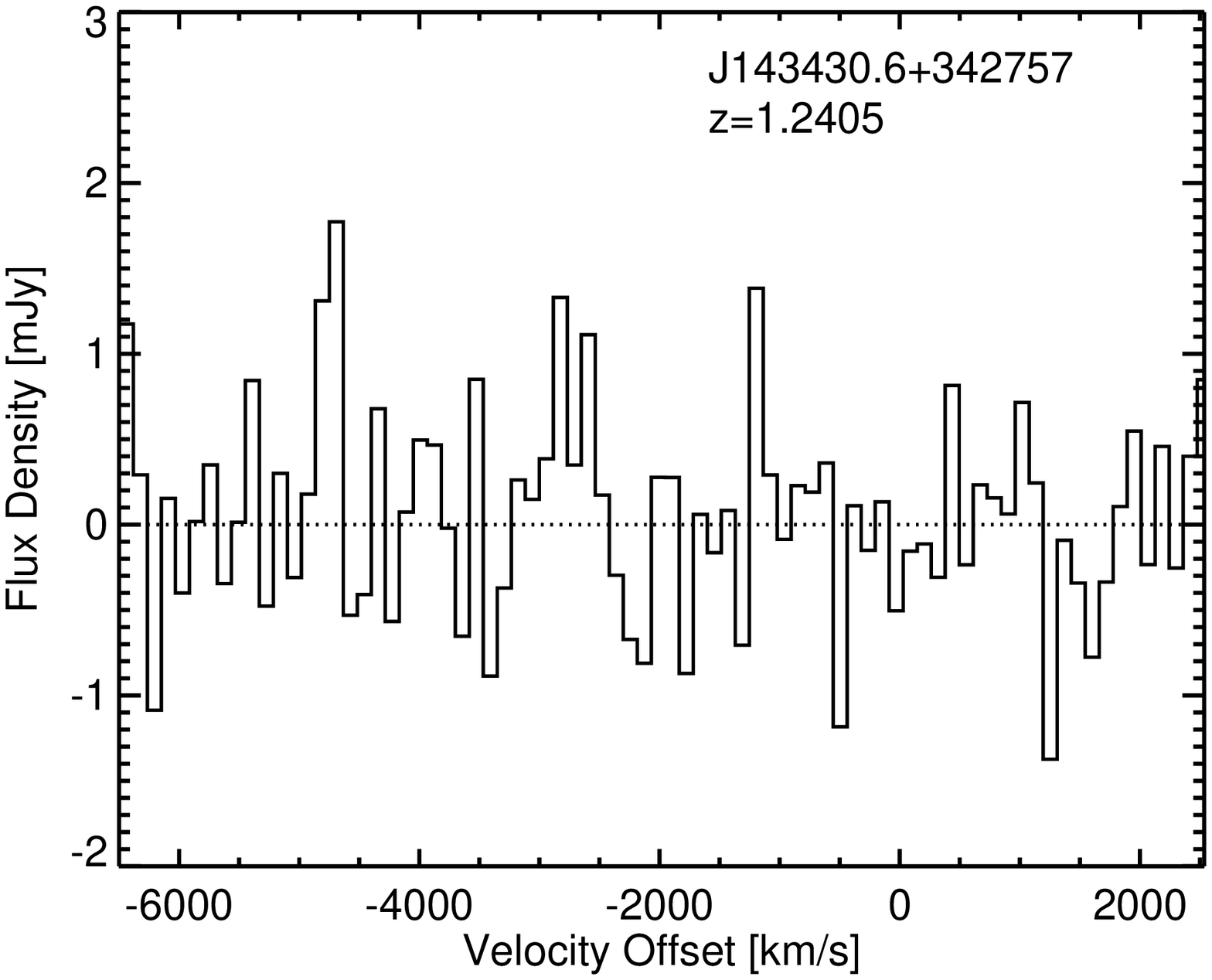,width=3.in}
\caption{PdBI 3~mm band spectra at the expected redshifts of \cotwo line emission in DOG1 (\textit{top}) and DOG2 (\textit{bottom}). Velocity offsets are relative to $z = 1.1160$ for DOG1 and $z = 1.2405$ for DOG2. The spectra have been resampled to 40.1~MHz resolution, corresponding to 110.1 and 116.5~km~s$^{-1}$. At this spectral resolution, the rms is 0.67 and 0.57~mJy, respectively.} 
\label{fig:dogspectra}
\end{figure}

We do not detect \cotwo line emission or 3~mm continuum emission in either of the DOGs (Figure~\ref{fig:dogspectra}). 
Table~\ref{tab:tab1} lists the rms sensitivity of the spectra, calculated for a channel width of 100~km~s$^{-1}$. We can calculate upper limits to the \cotwo line luminosity following the definition given by Solomon \& Vanden~Bout (2005):
\begin{equation}
L^\prime_{\rm CO} = 3.25 \times 10^7 \: S_{\rm CO} \: \Delta V \: \nu^{-2}_{\rm obs}\:  D_L^2 \: (1+z)^{-3} ~\rm [K~km~s^{-1}~pc^2]\:\: ,
\end{equation}
where $D_L$ is the luminosity distance (Mpc), $\nu_{\rm obs}$  is the frequency of observation (GHz), and in the case of a non-detection, $S_{\rm CO} \: \Delta V$ is the 3-$\sigma$ upper limit to the integrated intensity (Jy~km~s$^{-1}$), defined as 3$\cdot \sqrt{\Delta V_{FWHM} / \Delta V_{chan}} \cdot \sigma_{chan} \Delta V_{chan}$. The rms in a channel of width $\Delta V_{chan}$  (km~s$^{-1}$) is given by $\sigma_{chan}$ (Jy), and $\Delta V_{FWHM}$ (km~s$^{-1}$) is the assumed linewidth. For DOG1 and DOG2, we calculate 3-$\sigma$ upper limits to the \cotwo line luminosity, $L^\prime_{\rm CO} <  6.08\times 10^{9}$~K~km~s$^{-1}$~pc$^2$ and 
$< 6.63\times 10^{9}$~K~km~s$^{-1}$~pc$^2$, respectively, assuming a CO linewidth  of 300~km~s$^{-1}$, typical of 
luminous infrared quasar host galaxies at high-redshift (e.g. Carilli \& Wang 2006). The linewidth assumed for 
these calculations is narrower than that observed in high-redshift submm galaxies (SMGs, $\sim$800~km~s$^{-1}$; Greve et al.\ 2005), and we note that our upper limits to the line luminosity would increase by $\sim$60\% if we were to assume the broader CO linewidth.

In order to convert the limits on the \cotwo line luminosities to molecular gas mass estimates, we first assume that the \co and \cotwo lines are in thermal equilibrium, so that the luminosity in these two transitions is the same. This is the case for BzK selected star-forming galaxies at $z \sim 1.5$ (Daddi et al.\ 2008; Dannerbauer et al.\ 2009; Aravena et al.\ 2010). To convert the CO line luminosity to a molecular Hydrogen gas mass estimate, we adopt the conversion factor determined for nearby ULIRGs,  $\alpha_{\rm CO} \sim$$0.8 \: M_{\odot}$~(K~\kms~pc$^2$)$^{-1}$ (Downes \& Solomon 1998), as this is also the value generally 
assumed for FIR-luminous, high-redshift SMGs and quasar host galaxies. We note that the conversion factor is believed to be $\sim$5$\times$ greater for quiescent galaxies like the Milky Way (Solomon \& Barrett 1991), but the presence of a mid-infrared luminous AGN in the DOGs would favour the lower value. Adopting this conversion factor yields 3-$\sigma$ upper limits to the molecular gas mass, $M_{\rm H_2} < 4.86 \times 10^{9} \:$~M$_{\odot}$ (DOG1) and  $M_{\rm H_2} < 5.30 \times 10^{9} \:$~M$_{\odot}$ (DOG2). The estimated molecular gas masses are consistent with what is observed in ULIRGs in the local Universe (e.g. Downes \& Solomon 1998), but roughly an order of magnitude lower than in high-redshift SMGs (e.g. Greve et al.\ 2005).

\subsection{Serendipitous detection of \cotwo line emission in a cluster member}
\label{sec:comember}

\begin{figure}[ht]
\centering

\epsfig{file=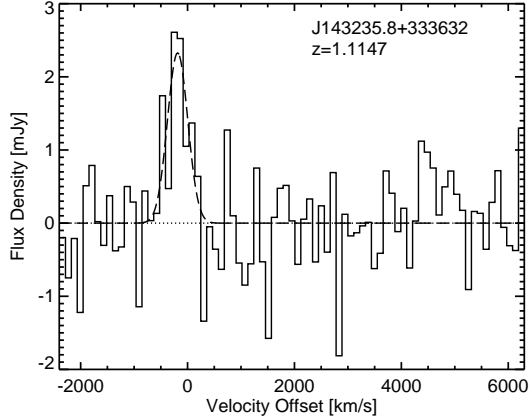,width=3.in}
\caption{PdBI 3~mm band spectrum of \cotwo line emission in the source serendipitously detected in the DOG1 image, referred  to here as COserendip. Velocity offsets are relative to $z = 1.1160$ (the optical redshift of DOG1), and the spectrum has been resampled to 40.1~MHz resolution, corresponding to 110.1~km~s$^{-1}$. At this spectral resolution the rms is 0.63~mJy. The CO redshift for this object is $z_{\rm co} = 1.1147$. The \textit{dashed line} shows the best-fit Gaussian profile with peak flux density, $S_p = 2.3\pm 0.7$~mJy,  and linewidth, $\Delta V_{FWHM} = 447 \pm 158$~km~s$^{-1}$.} 
\label{fig:serenspectra}
\end{figure}

\begin{figure}[ht]
\centering
\epsfig{file=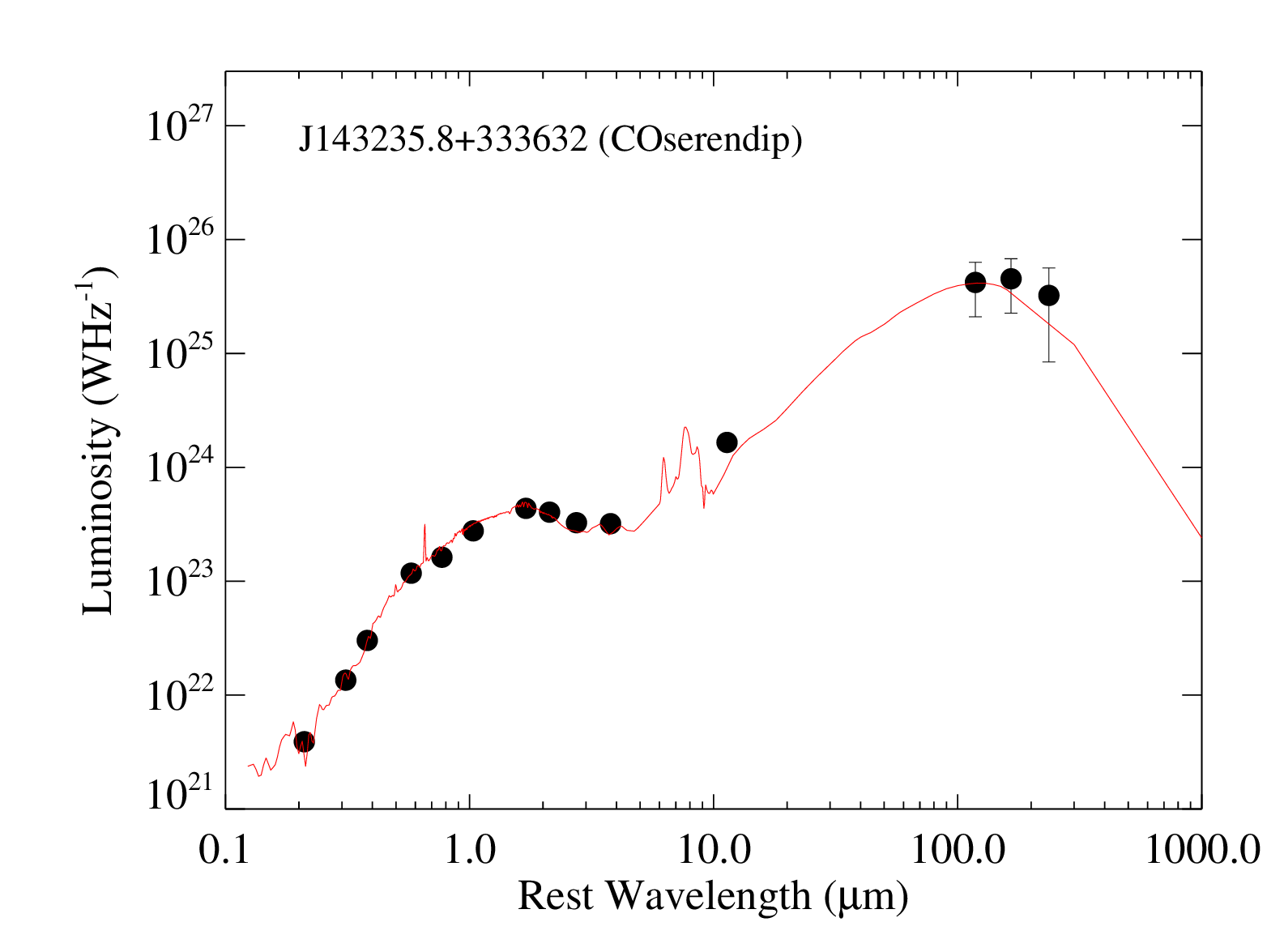,width=3.in}

\epsfig{file=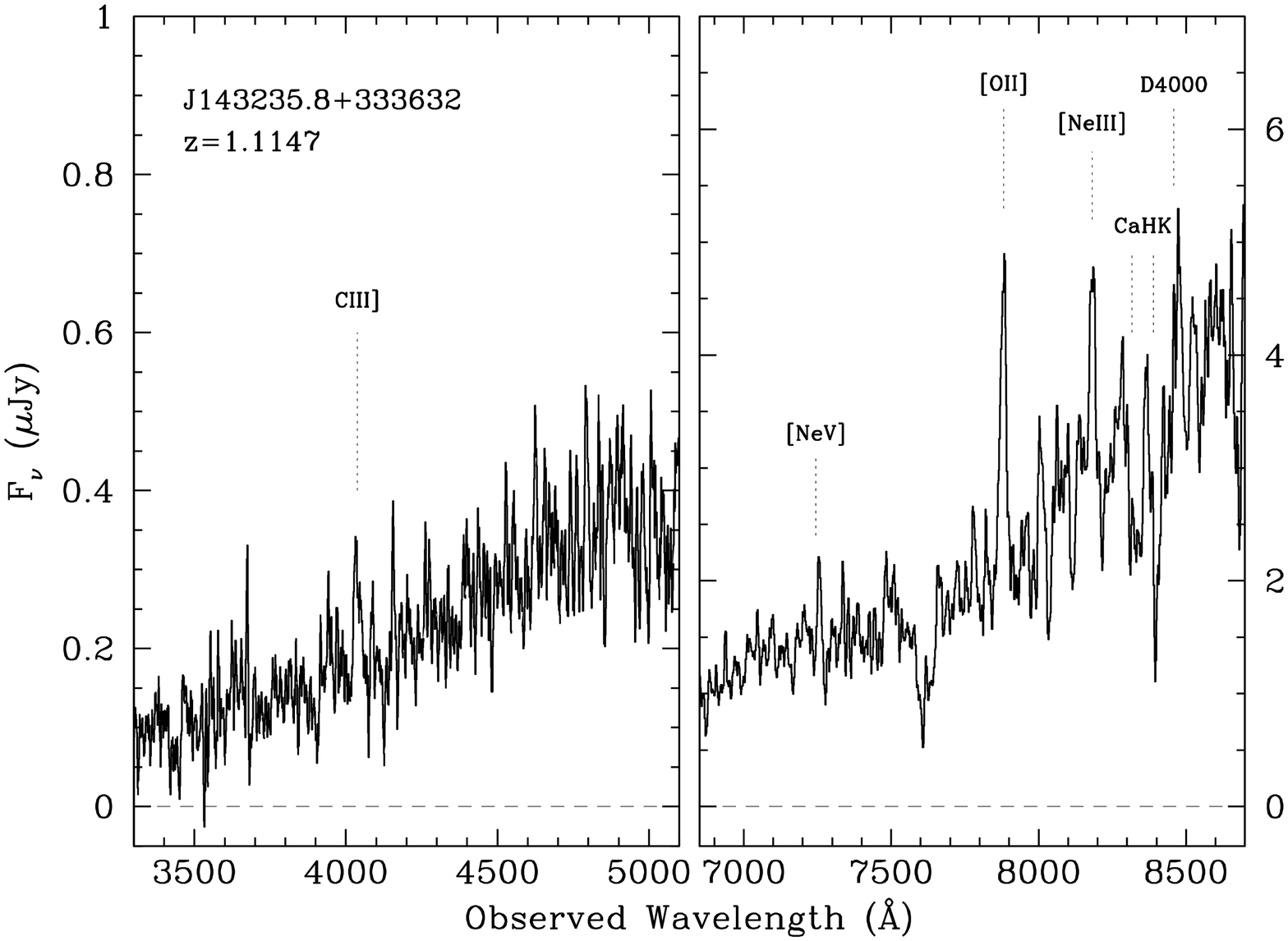,height=3.2in,angle=270}
\caption{\textit{top:}  Optical through submm-wavelength photometry of J143235.8+333632 (COserendip) at $z = 1.1147$. The template spectral energy distribution is that of a redshifted Seyfert 2 galaxy from the SWIRE templates of Polletta et al.\ (2007).
\textit{bottom:} Optical spectrum of COserendip, obtained with Keck/LRIS on 
UT 2005 June 3.  The position of this target on the slitmask left
a large, $\sim 2000$~\AA\, gap between the blue and red arms of the
dual-beam spectrograph, plotted separately above.  Observations
totaled 2.5~hr in clear conditions and 0\farcs9 seeing.  The spectrum
shows a clear detection of narrow \ion{C}{3}] emission, as well as 
strong [\ion{O}{2}] and [\ion{Ne}{3}], suggesting that this is a type-2 AGN
in the galaxy cluster.  The spectrum also shows more typical features
of evolved, early-type galaxies, including Ca~II~H$+$K absorption and
the D4000 continuum break. The [NeV] line emission is tentatively detected.} 
\label{fig:serendipsed}
\end{figure}

Given that the DOGs reside in clusters, and the large bandwidth (3.6~GHz) and field of view ($\sim$46$''$ diameter) covered by these observations, it is possible to search the spectral line data cubes for \cotwo line emission from other cluster members. We detect one object in \cotwo line emission, $\sim$12$''$ to the South-East of DOG1, which we hereafter refer to as SST24~J143235.8+333632, or `COserendip' (Figure~\ref{fig:serenspectra}). At this position, there is a cluster member with a measured optical spectroscopic redshift of $z_{\rm opt} = 1.114 \pm 0.001$, determined from the [\ion{O}{2}] line (Figure~\ref{fig:serendipsed}). The presence of \ion{C}{3}] line emission, as well as the tentative detection of [NeV], suggests a type-2 AGN, while a high [\ion{Ne}{3}]-to-[\ion{O}{2}] intensity ratio also supports the classification of COserendip as an obscured AGN (e.g. Villar-Mart{\'{\i}}n et al.\ 2008). 
The \cotwo line emission detected in this object is offset from the optical redshift of DOG1 by approximately $-200$~km~s$^{-1}$, corresponding to a redshift of $z_{\rm co} = 1.1147$, in good agreement with the redshift of the optical counterpart. The integrated intensity of this line is $S_{\rm CO} \: \Delta V = 1.16 \pm 0.21$~Jy~km~s$^{-1}$. Although the line emission is formally detected at 5.5~$\sigma$, the parameters of a Gaussian profile fit to the line are not well constrained, with a best-fit peak flux density, $S_p = 2.3\pm 0.7$~mJy, and line full width at half maximum, $\Delta V_{FWHM} = 447 \pm 158$~km~s$^{-1}$.  We estimate the infrared and FIR luminosities for COserendip using the best-fit spectral energy distribution (Figure~\ref{fig:serendipsed}), which is that of a Seyfert 2 galaxy taken from the SWIRE templates of Polletta et al.\ (2007). 
 The broad width of the CO line could be an indication that the molecular gas is rotating in a massive circumnuclear disk. Such broad linewidths are common in the high-redshift SMG population (e.g. Greve et al.\ 2005), and the FIR-luminous quasar host galaxy population at $z \sim 6$ (Wang et al.\ 2010).

The integrated intensity of the \cotwo line emission in COserendip can be converted to a CO line luminosity following equation~1, so that $L^\prime_{\rm CO} =  (1.94 \pm 0.35)\times 10^{10}$~K~km~s$^{-1}$~pc$^2$. Following Kennicutt (1998), the infared luminosity in COserendip implies a star-formation rate, $SFR \le $150~M$_\odot$~yr$^{-1}$, which is an upper limit due to the unknown contribution from the AGN. If the CO line emission were associated with a circumnuclear disk, then it would argue that the molecular gas clouds are not likely in a virialized state, motivating the adoption of  the ULIRG conversion factor of Downes \& Solomon (1998) to estimate the total molecular gas mass. Assuming thermalized \cotwo line emission, we estimate, $M_{\rm H_2} = (1.55 \pm 0.28)\times 10^{10}$~M$_{\odot}$, which is comparable to the typical mass of gas in bright, blank-field SMGs and FIR-luminous quasar host galaxies at high-redshift.

\section{Discussion}
\label{sec:discuss}

The dust heating which gives rise to the infrared luminosities in DOG1 and DOG2 has a contribution from AGN activity, as indicated by the SED template fits showing that the mid-infrared emission in both objects can be described by a power-law. Although the AGN can boost the total infrared luminosity, the scatter in the $L_{\rm IR}$ to $L'_{\rm CO}$ ratio for nearby LIRGs and ULIRGs is large, even between those galaxies which do not contain a luminous AGN (e.g. Solomon et al.\ 1997).
 Considering the longer wavelength FIR emission, our $3-\sigma$ upper limits to $L_{FIR}$ are also in broad agreement with the limits on the \cotwo line luminosities given the observed scatter in the $L_{\rm FIR}-L'_{\rm CO}$ relation for nearby and high-redshift ULIRGs and more quiescent star-forming galaxies, independent of whether or not infrared luminous quasars are included in this relation (e.g. Solomon \& Vanden~Bout 2005; Riechers et al.\ 2006; Daddi et al.\ 2008). As such, it is not clear whether the molecular gas content of these two DOGs found at the outskirts of massive clusters have been affected by the cluster environment. More sensitive observations of redshifted CO line emission with the Atacama Large Millimeter Array (ALMA) would determine if these two objects deviate from the  $L_{\rm FIR}-L'_{\rm CO}$ relation.

The ratio of the FIR luminosity to the total molecular gas mass in a galaxy provides a measure of its star formation efficiency ($SFE = L_{FIR} / M_{\rm H_2}$). In high-redshift FIR-luminous quasar host galaxies, the average star formation efficiency is, $SFE \sim $430~L$_\odot$~M$_{\odot}^{-1}$ (e.g. Solomon \& Vanden~Bout 2005). For ULIRGs in the local Universe, the star formation efficiency is generally a factor of $\sim$4$\times$ higher than in more quiescent galaxies like the Milky Way (e.g. Bouch\'e et al.\ 2007), and consistent with that in high-redshift SMGs ($\ga$100~L$_\odot$~M$_{\odot}^{-1}$; e.g. Frayer et al.\ 2011). 
 In the case of DOG1 and DOG2, we have only upper limits to the FIR and CO line luminosities, and so are unable to estimate meaningful star-formation efficiencies for these objects.
 In the case of COserendip, the SED template fitting indicates $L_{\rm FIR} < 4.9\times 10^{11}$~L$_{\odot}$, and so we estimate that the star formation efficiency is $SFE < 32$~L$_{\odot}$~M$_{\odot}^{-1}$, a value typical of that observed in nearby luminous infrared galaxies (e.g. Gao \& Solomon 2004), but not as high as that observed in FIR luminous, high-redshift quasar host galaxies. The star formation efficiency estimated for COserendip is also similar to that observed in two star-forming galaxies detected in \co line emission in the outskirts of a rich galaxy cluster at $z \sim 0.4$ (Geach et al.\ 2009).

The specific $SFR$ ($sSFR  = SFR/ M_*$)  is shown to follow a well-defined redshift evolution for typical star-forming galaxies, while more extreme starburst galaxies fall above this relation (e.g. Brinchmann et al.\ 2004; Pannella et al.\ 2009; Elbaz et al.\ 2011). In the case of the DOGs, the significant uncertainties on their star formation rates prevent us from estimating stringent limits on the value of $sSFR$. 
For COSerendip, the estimated stellar mass and our upper limit to the $SFR$, implies $sSFR \le 1.6$~[Gyr]$^{-1}$. This value is $\sim$3$\times$ larger than the tight relation defined by the main-sequence of star-forming galaxies at $z \sim 1.1$ (Elbaz et al.\ 2011), potentially placing COserendip among starburst galaxies. 
Assuming that the upper limit to the star formation rate in COserendip ($\sim$150~M$_{\odot}$~yr$^{-1}$) could be maintained, its estimated molecular gas mass would be exhausted in $\sim$100~Myr, and the stellar mass would grow to $\sim$1.1$\times$10$^{11}$~M$_{\odot}$. 

All of our targets show some evidence for AGN activity and are located outside the virial radii of the two clusters. Studies of the long-wavelength radio, X-ray and thermal dust emission in active galaxies associated with massive clusters at lower redshifts have found that the number density of AGN within the central regions of clusters increases with redshift (e.g. Galametz et al.\ 2009; Martini et al.\ 2009). Little is known about the distribution in individual objects of the molecular gas fueling this actvity in intermediate redshift ($z \sim 0.5$) cluster galaxies, and previous observations of \loj CO line emission have targetted star-forming galaxies in the process of cluster infall (Geach et al.\ 2009).  The only exception to this is a recent detection of CO~\textit{J=}3-2 line emission in a spiral galaxy found inside the virial radius of the cluster Abell~370 (Kanekar et al.\ \textit{in prep.}). However, this galaxy does not appear to contain an AGN, and so our detection of CO line emission in COserendip at $z \sim 1$ provides the only clear evidence for cold molecular gas in infrared luminous AGN associated with  overdense cluster environments.

\section{Summary}
\label{sec:sum}

We present the first observations of cold molecular gas, as traced by \cotwo line emission, in members of massive ($>10^{14}$~M$_{\odot}$) galaxy clusters at $z \sim 1$. Despite their significant estimated infrared luminosities, two ULIRGs are not detected in CO line emission, implying cold molecular gas masses $M_{\rm H_2} \la 5\times10^9$~M$_{\odot}$.  Given the presence of AGN in these DOGs, which contributes to the dust heating, estimates of their star formation efficiencies are highly uncertain, and our non-detections of CO line emission are not sensitive enough to determine if these objects deviate from the $L_{\rm FIR}-L'_{\rm CO}$ relation. We detect \cotwo line emission in a third cluster member, implying a molecular gas mass of $M_{\rm H_2} = (1.55 \pm 0.28)\times 10^{10}$~M$_{\odot}$. Contrary to the local Universe where galaxy clusters are passively evolving, our detection of CO line emission in COserendip provides the first evidence for the existence of cold molecular gas in a $z \sim 1$ cluster member, demonstrating that fuel can be available in the outskirts of clusters to drive active star formation or AGN activity.

\section{Acknowledgments}

We thank Jan-Martin Winters, Melanie Krips and Roberto Neri for their helpful advice and guidance with the data analysis. For their contributions to this survey, we thank Anthony Gonzalez and Thomas Soifer, as well as Adam Stanford for the LRIS spectrum. This work was co-funded under the Marie Curie Actions of the European Commission (FP7-COFUND). Some of the data presented herein were obtained at the W.M. Keck Observatory, which is operated as a scientific partnership among the California Institute of Technology, the University of California and the National Aeronautics and Space Administration. The Observatory was made possible by the generous financial support of the W.M. Keck Foundation. Part of this work is based on observations made with the {\it Spitzer Space Telescope}, which is operated by the Jet Propulsion Laboratory, California Institute of Technology under a contract with NASA. Support for this work was provided by NASA. {\it Herschel} is an ESA space observatory with science instruments provided by European-led Principal Investigator consortia and with important participation from NASA.

\end{document}